\documentclass[smallextended]{svjour3}
\usepackage{graphicx, amsmath, amssymb}
\usepackage[caption=false]{subfig}

\newcommand{\braket}[2]{{\left\langle #1 \middle| #2 \right\rangle}}
\newcommand{\bra}[1]{{\left\langle #1 \right|}}
\newcommand{\ket}[1]{{\left| #1 \right\rangle}}
\newcommand{\ketbra}[2]{{\left| #1 \middle\rangle \middle \langle #2 \right|}}

\journalname{Quantum Inf Process}

\begin{document}

\title{Exceptional Quantum Walk Search on the Cycle}

\author{Thomas G.~Wong \and \\ Raqueline A.~M.~Santos}

\authorrunning{T.~G.~Wong \and R.~A.~M.~Santos}

\institute{T.~G.~Wong \at
	   Department of Computer Science, University of Texas at Austin, 2317 Speedway, Austin, Texas 78712, USA \\
	   \email{twong@cs.utexas.edu}
	   \and
	   R.~A.~M.~Santos \at
	   Center for Quantum Computer Science, University of Latvia, Rai\c{n}a bulv.~19, R\=\i ga, LV-1586, Latvia \\
	   \email{rsantos@lu.lv}
}

\date{Received: date / Accepted: date}

\maketitle

\begin{abstract}
	Quantum walks are standard tools for searching graphs for marked vertices, and they often yield quadratic speedups over a classical random walk's hitting time. In some exceptional cases, however, the system only evolves by sign flips, staying in a uniform probability distribution for all time. We prove that the one-dimensional periodic lattice or cycle with any arrangement of marked vertices is such an exceptional configuration. Using this discovery, we construct a search problem where the quantum walk's random sampling yields an arbitrary speedup in query complexity over the classical random walk's hitting time. In this context, however, the mixing time to prepare the initial uniform state is a more suitable comparison than the hitting time, and then the speedup is roughly quadratic.
	\keywords{Quantum walk \and Quantum search \and Spatial search \and Exceptional configuration \and Random walk \and Markov chain \and Hitting time \and Mixing time}
	\PACS{03.67.Ac, 02.10.Ox, 02.50.Ga, 05.40.Fb}
\end{abstract}


\section{Introduction}

\begin{figure}
\begin{center}
	\subfloat[]{
		\includegraphics{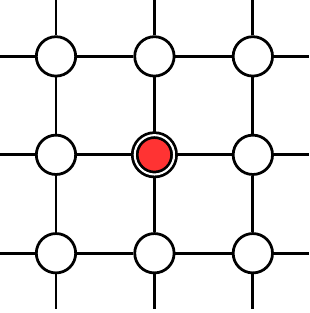}
		\label{fig:grid3_onemarked} 
	} \quad \quad \quad \quad \quad \quad
	\subfloat[]{
		\includegraphics{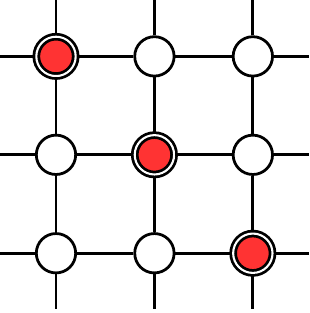}
		\label{fig:grid3_diagonal}
	}
	\caption{Periodic square lattice of $N = 3 \times 3$ vertices with a (a) single marked vertex and (b) marked diagonal. Marked vertices are denoted by double circles and colored red.}
\end{center}
\end{figure}

The ability to search for information, such as in a database or on a network, is one of the fundamental problems in information science. This can be modeled as search on a graph for a marked vertex, such as the two-dimensional (2D) periodic square lattice (or discrete torus or grid) depicted in Fig.~\ref{fig:grid3_onemarked} with a unique marked vertex. One approach to finding the marked vertex is to start at a random vertex and then randomly walk around on the graph until the marked vertex is found. The expected number of steps this takes is defined as the \emph{hitting time}, and for the 2D grid with $N$ vertices, it is $O(N \log N)$. Note this is worse than simply guessing for the marked vertex, which takes an expected $O(N)$ guesses, because the random walk is restricted to local movements on the graph and may revisit vertices.

This local search procedure can be made quantum mechanical by replacing the classical random walk with its quantum analogue, the quantum walk. There are several definitions of quantum walks, the best of which search the 2D grid for a unique marked vertex in $O(\sqrt{N \log N})$ time \cite{Tulsi2008,Magniez2012,Ambainis2013}, which is a quadratic speedup over the classical random walk's hitting time. This assumes that the quantum walk search algorithm may need to be repeated since the probability of measuring the quantum walker at a marked vertex may be less than $1$, in contrast to the classical random walk which will eventually find a marked vertex with certainty.

Such quadratic speedups over hitting time are achievable under some circumstances, and much research has been done to explore when they occur. Magniez \textit{et al.}~\cite{Magniez2012} gave a quantum walk that achieves a quadratic speedup over hitting time when the classical Markov chain is state-transitive, reversible, and ergodic. Further improvements were made by Krovi \textit{et al.} \cite{Krovi2016}, who gave a quantum walk that achieves the quadratic speedup without the requirement that the classical Markov chain be state-transitive. In general, it is unknown if quantum walks can always achieve a quadratic speedup over the classical random walk's hitting time, and it is also unknown if a greater-than-quadratic speedup is attainable.

One might also have a search problem with multiple marked vertices, where the goal is to find any one of them. Krovi \textit{et al.} \cite{Krovi2016} also investigated this, giving a quantum walk that searches quadratically faster than a quantity called the ``extended'' hitting time, which is equivalent to the hitting time with one marked vertex and lower-bounded by it with multiple marked vertices. Ambainis and Kokainis \cite{AK2015}, however, showed that the separation between the extended and usual hitting times can be asymptotically infinite.

Classically, having additional marked vertices makes the search problem easier because there are more marked vertices to randomly walk into. Quantumly, however, the precise location of the marked vertices, such as whether they are grouped together or spread apart, can affect the performance of the quantum walk search algorithm either favorably or unfavorably.

For example, consider the discrete-time coined quantum walk that was first proposed by Meyer in the context of quantum cellular automata \cite{Meyer1996a,Meyer1996b} and later recast as a quantum walk by Aharonov \textit{et al.}~\cite{Aharonov2001}. Its ability to search the 2D grid has been explored for several arrangements of marked vertices \cite{AKR2005,AR2008,NR2016a,NR2016b,NS2016,Wong24}, and one problematic configuration is when the marked vertices lie along a diagonal, such as in Fig.~\ref{fig:grid3_diagonal}. In this case, the coined quantum walk search algorithm, which begins in an equal superposition over all the vertices, only evolves by acquiring minus signs \cite{AR2008}. Then the system remains in a uniform probability distribution for all time, which means the quantum algorithm is equivalent to classically guessing for a marked vertex. Such arrangements of marked vertices, where the quantum algorithm is no better than classically guessing, are called \emph{exceptional configurations} \cite{AR2008,NR2016b}.

\begin{figure}
\begin{center}
	\includegraphics{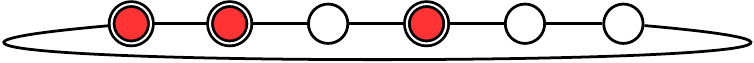}
	\caption{\label{fig:line_N6_marked_unlabeled} A cycle of $N = 6$ vertices, $k = 3$ of which are marked (denoted by double circles and colored red).}
\end{center}
\end{figure}

In this paper, rather than using coined quantum walks, we work in the equivalent framework of Szegedy's quantum walk \cite{Szegedy2004,Magniez2012,Wong26}, which is a direct quantization of a classical random walk or Markov chain. We prove that \emph{any} configuration of marked vertices on the 1D cycle, an example of which is shown in Fig.~\ref{fig:line_N6_marked_unlabeled}, is an exceptional configuration for this walk. That is, for the cycle, the quantum walk only evolves by sign flips, and so it stays in its initial uniform probability distribution and is equivalent to classically guessing and checking. We utilize this observation to construct a search problem with an arbitrary separation in query complexity between the quantum walk and the classical random walk's hitting time, hence demonstrating a greater-than-quadratic speedup by quantum walk. This speedup is somewhat artificial, however, relying on the quantum walk's ability to sample uniformly rather than evolve nontrivially. As such, the mixing time to prepare the initial state is a more suitable comparision, and then the quantum walk achieves a nearly quadratic speedup over the classical random walk. We end by observing that any higher-dimensional graph that reduces to the 1D line is also an exceptional configuration.


\section{Quantum Walk on the Cycle}

\begin{figure}
\begin{center}
	\subfloat[]{
		\includegraphics{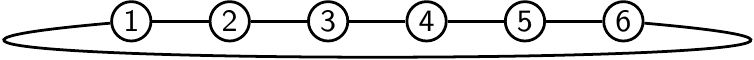}
		\label{fig:line_N6} 
	} \quad \quad \quad \quad \quad \quad
	\subfloat[]{
		\includegraphics{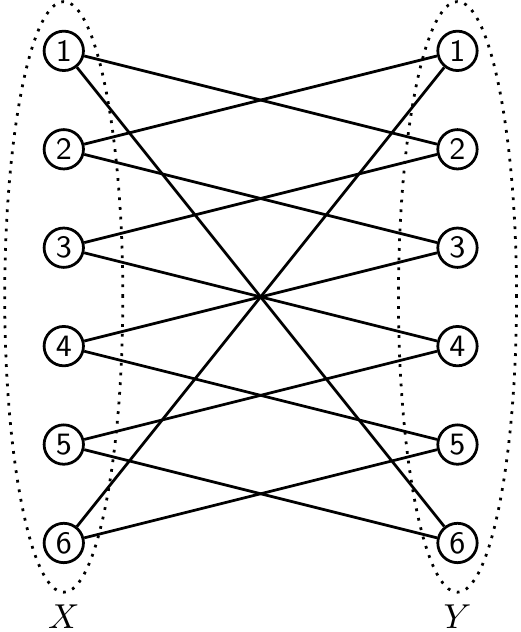}
		\label{fig:line_N6_double} 
	}
	\caption{(a) The 1D cycle of $N = 6$ vertices and (b) its bipartite double cover for Szegedy's quantum walk.}
\end{center}
\end{figure}

We begin by introducing Szegedy's quantum walk \cite{Szegedy2004} without searching. For concreteness, we consider a small example of the 1D cycle with $N = 6$ vertices, which is depicted in Fig.~\ref{fig:line_N6}. Our analysis generalizes to arbitrary $N$ in a straightforward manner. For a textbook introduction, see \cite{Portugal2013}. 

Szegedy's quantum walk is a quantization of a classical Markov chain, so we start with a classical random walk on the cycle. In Fig.~\ref{fig:line_N6}, we labeled the vertices $1, 2, \dots, 6$. Using these as a basis and treating each edge equally, the stochastic transition matrix $P$ of the classical random walk is
\[ P = \frac{1}{2} \begin{pmatrix}
	0 & 1 & 0 & 0 & 0 & 1 \\
	1 & 0 & 1 & 0 & 0 & 0 \\
	0 & 1 & 0 & 1 & 0 & 0 \\
	0 & 0 & 1 & 0 & 1 & 0 \\
	0 & 0 & 0 & 1 & 0 & 1 \\
	1 & 0 & 0 & 0 & 1 & 0 \\
\end{pmatrix}, \]
where $P$ acts on states on the right, so $P_{ij}$ is the probability of transitioning from vertex $i$ to $j$. For example, a walker at vertex $1$ has probability $1/2$ of jumping to each of the vertices $2$ and $6$.

To turn this Markov chain into Szegedy's quantum walk, we first construct the bipartite double cover of the original graph. For the 1D cycle in Fig.~\ref{fig:line_N6}, its bipartite double cover is shown in Fig.~\ref{fig:line_N6_double}, and to explain this, let us call the partite sets of Fig.~\ref{fig:line_N6_double} $X$ and $Y$. Each of these sets gets a copy of the vertices in the original cycle. A vertex $x \in X$ is adjacent to a vertex $y \in Y$ if and only if $x$ and $y$ are adjacent in the original graph. For example, in the 1D cycle in Fig.~\ref{fig:line_N6}, vertex $1$ is adjacent to vertices $2$ and $6$. Then in its bipartite double cover in Fig.~\ref{fig:line_N6_double}, vertex $1 \in X$ is adjacent to vertices $2,6 \in Y$. Conversely, vertex $1 \in Y$ is adjacent to  vertices $2,6 \in X$. Formally, if the original graph is $G$, its bipartite double cover is $G \times K_2$, where $K_2$ is the complete graph of $2$ vertices. 

Szegedy's quantum walk occurs on the edges of the bipartite double cover. This is because discrete-time quantum walks require additional degrees of freedom \cite{Meyer1996a,Meyer1996b} or staggered quantum operations \cite{Portugal2016} to evolve nontrivially. Since the edges connect vertices in $X$ with vertices in $Y$, the edges are spanned by the following orthonormal computational basis:
\[ \{ \ket{x,y} : x \in X, y \in Y \}, \]
where $\ket{x,y}$ denotes $\ket{x} \otimes \ket{y}$. Thus, the Hilbert space is $\mathbb{C}^N \otimes \mathbb{C}^N$, where $N$ is the number of vertices in the original graph.

Now Szegedy's walk is defined by repeated applications of
\[ W_P = R_b R_a, \]
where
\[ R_a = 2 \sum_{x \in X} \ketbra{\phi_x}{\phi_x} - I \]
\[ R_b = 2 \sum_{y \in Y} \ketbra{\psi_y}{\psi_y} - I \]
are reflection operators defined by
\[ \ket{\phi_x} = \ket{x} \otimes \sum_{y \in Y} \sqrt{P_{xy}} \ket{y} \]
\[ \ket{\psi_y} = \sum_{y \in Y} \sqrt{P_{yx}} \ket{x} \otimes \ket{y}. \]
Let us analyze what these operators do. First note that $P_{xy} = 1 / \text{deg}(x)$ when $x$ and $y$ are adjacent, since the original graph is unweighted. Specifically for the 1D cycle, $\deg(x) = 2$, but we leave it as $\deg(x)$ in the subsequent calculations for generality to other graphs. Then
\[ \ket{\phi_x} = \ket{x} \otimes \frac{1}{\sqrt{\text{deg}(x)}} \sum_{y \sim x} \ket{y} = \frac{1}{\sqrt{\text{deg}(x)}} \sum_{y \sim x} \ket{x,y} . \]
So $\ket{\phi_x}$ is the equal superposition of edges from vertex $x \in X$. Inserting this into the reflection $R_a$, we get
\begin{align*}
	R_a 
	&= 2 \sum_{x \in X} \frac{1}{\sqrt{\text{deg}(x)}} \sum_{y \sim x} \ket{x,y} \frac{1}{\sqrt{\text{deg}(x)}} \sum_{y' \sim x} \bra{x,y'} - I \\
	&= 2 \sum_{x \in X} \frac{1}{\text{deg}(x)} \sum_{y \sim x} \ket{x,y} \sum_{y' \sim x} \bra{x,y'} - I.
\end{align*}
To understand what $R_a$ does, let us apply it to an arbitrary state
\[ \sum_{x \in X} \sum_{y \sim x} c_{x,y} \ket{x,y}, \]
where $c_{x,y}$ is the amplitude of the state on the edge joining vertices $x$ and $y$. $R_a$ acts on this by
\[ R_a \sum_{x \in X} \sum_{y \sim x} c_{x,y} \ket{x,y} = 2 \sum_{x \in X} \frac{1}{\text{deg}(x)} \sum_{y \sim x} \ket{x,y} \sum_{y' \sim x} c_{x,y'} - \sum_{x \in X} \sum_{y \sim x} c_{x,y} \ket{x,y}. \]
But
\[ \bar{c}_x = \frac{1}{\text{deg}(x)} \sum_{y' \sim x} c_{x,y'} \]
is the average amplitude of the edges from $x$. So we have
\begin{align*}
	R_a \sum_{x \in X} \sum_{y \sim x} c_{x,y} \ket{x,y} 
	&= 2 \sum_{x \in X} \bar{c}_x \sum_{y \sim x} \ket{x,y} - \sum_{x \in X} \sum_{y \sim x} c_{x,y} \ket{x,y} \\
	&= \sum_{x \in X} \sum_{y \sim x} \left( 2 \bar{c}_x - c_{x,y} \right) \ket{x,y}.
\end{align*}
Note that $2 \bar{c}_x - c_{x,y}$ inverts the amplitude $c_{x,y}$ about the average $\bar{c}_x$. Thus, $R_a$ goes through each vertex $x \in X$ and inverts its edges about their average value. $R_b$ acts similarly, except it goes through $Y$. This observation that $R_a$ and $R_b$ invert about average amplitudes is likely known, but this seems to be the first time it has been explicitly shown in the literature. Importantly, it plays a central role in our proof that the system only evolves by flipping signs.

Finally, we measure the quantum walk in the $X$ partite set, so we take the partial trace over $Y$. For example, if the state of the system is $\ket{\psi}$, then the probability of measuring the walker at vertex $1 \in X$ is
\[ |\braket{1,2}{\psi}|^2 + |\braket{1,6}{\psi}|^2. \]


\section{Search on the Cycle}

\begin{figure}
\begin{center}
	\includegraphics{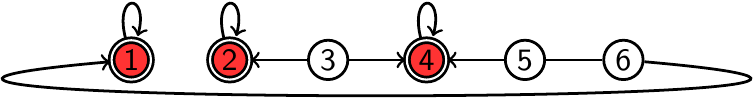}
	\caption{\label{fig:line_N6_marked} The 1D cycle of $N = 6$ vertices with $k = 3$ absorbing marked vertices, indicated by double circles and colored red.}
\end{center}
\end{figure}

Now let us consider searching using the example in Fig.~\ref{fig:line_N6_marked_unlabeled}, which contains all possible relations between marked and unmarked vertices: marked adjacent to marked, marked adjacent to unmarked, and unmarked adjacent to unmarked. Recall that the classical random walk begins at a random vertex, which is equivalent to beginning in the uniform probability distribution over the vertices. The walker jumps around and stops once a marked vertex is found, so we can interpret the marked vertices as absorbing vertices, as shown in Fig.~\ref{fig:line_N6_marked}. The stochastic transition matrix corresponding to this absorbing random walk is
\[ P' = \frac{1}{2} \begin{pmatrix}
	2 & 0 & 0 & 0 & 0 & 0 \\
	0 & 2 & 0 & 0 & 0 & 0 \\
	0 & 1 & 0 & 1 & 0 & 0 \\
	0 & 0 & 0 & 2 & 0 & 0 \\
	0 & 0 & 0 & 1 & 0 & 1 \\
	1 & 0 & 0 & 0 & 1 & 0 \\
\end{pmatrix}. \]

\begin{figure}
\begin{center}
	\subfloat[]{
		\includegraphics{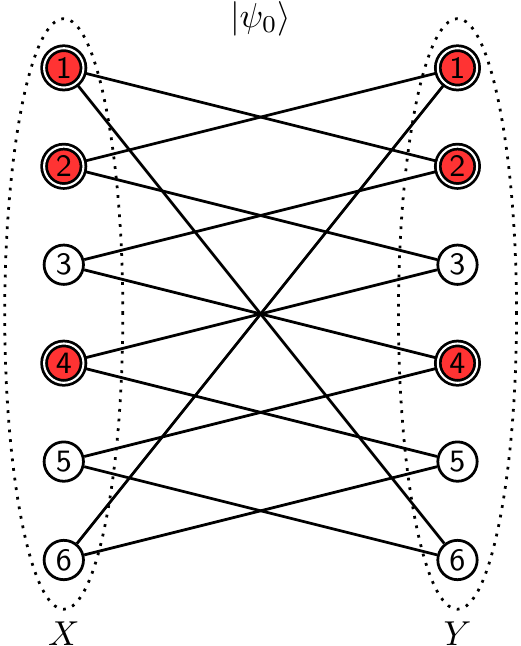}
		\label{fig:line_N6_double_marked_W0}
	} \quad
	\subfloat[]{
		\includegraphics{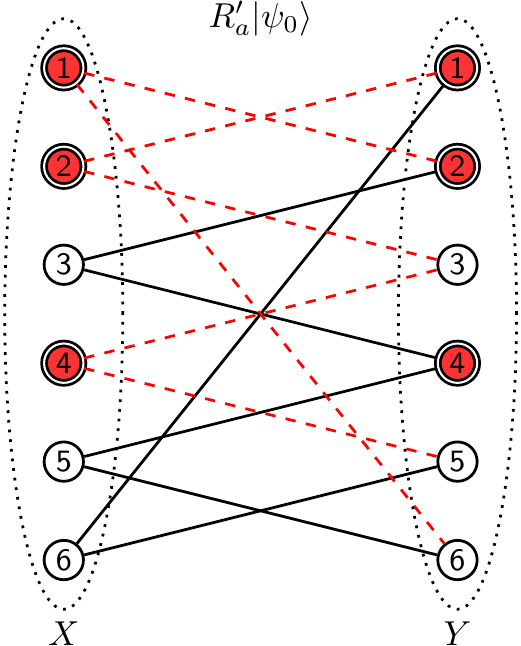}
		\label{fig:line_N6_double_marked_Ra} 
	}

	\subfloat[]{
		\includegraphics{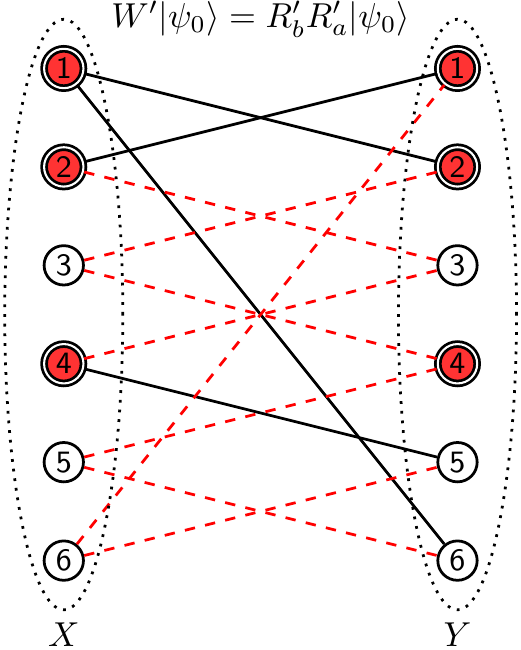}
		\label{fig:line_N6_double_marked_W1}
	} \quad
	\subfloat[]{
		\includegraphics{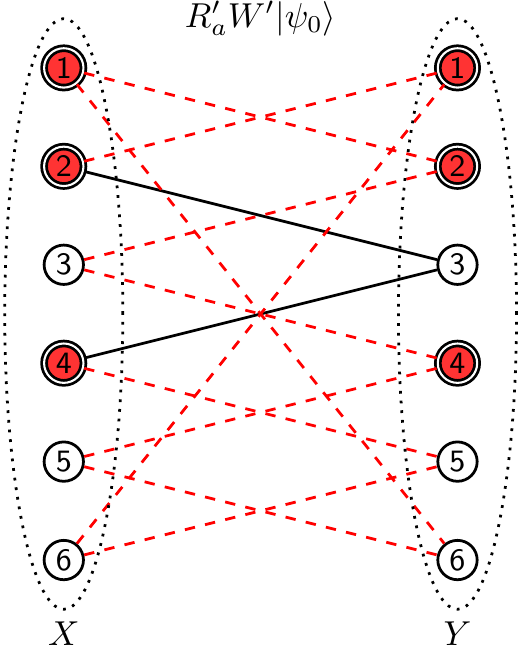}
		\label{fig:line_N6_double_marked_RaW1} 
	}
	\caption{Applications of Szegedy's quantum walk search operator. The solid black and dashed red edges respectively denote amplitudes of $\pm 1/\sqrt{12}$. Subfigure (a) is the initial state, (b) is after the first reflection, (c) is after the second reflection, which is one application of Szegedy's search operator, and (d) is after the third reflection.}
\end{center}
\end{figure}

To search using Szegedy's quantum walk, we begin in the state
\[ \ket{\psi_0} = \frac{1}{\sqrt{N}} \sum_{x,y} \sqrt{P_{xy}} \ket{x,y}, \]
which is defined in terms of $P$, not $P'$, since we do not yet know where the marked vertices are when initializing the state. For the cycle with $N = 6$ vertices, the initial state is depicted in Fig.~\ref{fig:line_N6_double_marked_W0}, where each edge has an amplitude of $1/\sqrt{12}$. That is, the state is a uniform superposition over the edges, in analogy to the classical random walk starting in a uniform probability distribution.

The system evolves using Szegedy's quantum walk
\[ W' = R_b' R_a', \]
where the primes denote that this uses the absorbing transition matrix $P'$. On unmarked vertices, this acts the same way as $W = R_b R_a$ that quantizes $P$, \textit{i.e.}, by inverting edges from a vertex about the average amplitude of edges from the vertex. The walk acts differently on marked vertices, however, as we now show.

Consider a marked vertex $i \in X$, such as vertex $1$, $2$, or $4$ in Fig.~\ref{fig:line_N6_double_marked_W0}. Then $\ket{\phi_{i}} = \ket{i,i}$ since $P'_{i,i} = 1$. Then the reflection operator $R_a'$ acts by on an edge joining $i \in X$ and $y \in Y$ by
\[ R_a' \ket{i,y} = 2 \ket{i,i} \braket{i,i}{i,y} - \ket{i,y} = \begin{cases}
	\ket{i,i}, & y = i \\
	-\ket{i,y}, & y \ne i
\end{cases}. \]
Since the amplitude at edge $\ket{i,i}$ is initially zero, it continues to be zero, and we can ignore it. All other edges incident to $i \in X$ acquire minus signs, and this inversion acts as a search oracle.

Now that all the operators are defined for search by Szegedy's quantum walk, let us apply it to our example. Again, the initial uniform state of the system $\ket{\psi_0}$ is depicted in Fig.~\ref{fig:line_N6_double_marked_W0}, where each edge has an amplitude of $1/\sqrt{12}$. Now we apply the first reflection operator $R_a'$. For marked vertices $1, 2, 4 \in X$, this flips the sign of incident edges. For unmarked vertices $3, 5, 6 \in X$, it inverts incident edges about their average at the vertex, and in this case, both edges at an unmarked vertex have amplitudes $1/\sqrt{12}$, so their average is $1/\sqrt{12}$, and the inversion does nothing. The resulting state is shown in Fig.~\ref{fig:line_N6_double_marked_Ra}, where a dashed red edge denotes an amplitude of $-1/\sqrt{12}$ (while a solid black edge is still positive $1/\sqrt{12}$).

Next we apply $R_b'$. For marked vertices in $1, 2, 4 \in Y$, this flips the sign of incident edges. For each unmarked vertex $3, 5, 6 \in Y$, edges are inverted about the average amplitude at the vertex. In particular, both edges incident to vertex $3 \in Y$ have amplitude $-1/\sqrt{12}$, so their average is $-1/\sqrt{12}$, and the inversion does nothing. On the other hand, vertex $5 \in Y$ has edges of amplitude $1/\sqrt{12}$ and $-1/\sqrt{12}$, so their average is zero, and both edges get flipped in the inversion about zero. This is similar for vertex $6 \in Y$. The result of applying $R_b'$ is shown in Fig.~\ref{fig:line_N6_double_marked_W1}, and this is one application of Szegedy's search operator $W' = R_b' R_a'$.

Applying $R_a'$ again, we flip the edges incident to marked vertices in $X$ and invert about the average edges incident to unmarked vertices, resulting in Fig.~\ref{fig:line_N6_double_marked_RaW1}. Continuing this, the system continues to only evolve by sign flips, and the signs are shown in Table~\ref{table:line_N6_signs}. Note that after a total of six applications of $W'$, the system returns to its initial state, \textit{i.e.}, $(W')^6 \ket{\psi_0} = (R_b' R_a')^6 \ket{\psi_0} = \ket{\psi_0}$.

\begin{table}
	\caption{\label{table:line_N6_signs} Amplitudes for search by Szegedy's quantum walk on the cycle of $N = 6$ vertices, where vertices $1$, $2$, and $4$ are marked, as depicted in Fig.~\ref{fig:line_N6_marked}. $\pm$ denotes amplitudes of $\pm 1/\sqrt{12}$.}
	\setlength{\tabcolsep}{5pt}
	\begin{center}
	\begin{tabular}{ccccccccc}
		\hline\noalign{\smallskip}
		\textbf{Edge} & $(W')^2$ & $R_a'(W')^2$ & $(W')^3$ & $R_a'(W')^3$ & $(W')^4$ & $R_a' (W')^4$ & $(W')^5$ & $R_a'(W')^5$ \\
		\noalign{\smallskip}\hline\noalign{\smallskip}
		$\ket{1,6}$ & $-$ & $+$ & $+$ & $-$ & $+$ & $-$ & $-$ & $+$ \\
		$\ket{1,2}$ & $+$ & $-$ & $+$ & $-$ & $+$ & $-$ & $+$ & $-$ \\
		$\ket{2,1}$ & $+$ & $-$ & $+$ & $-$ & $+$ & $-$ & $+$ & $-$ \\
		$\ket{2,3}$ & $+$ & $-$ & $-$ & $+$ & $+$ & $-$ & $-$ & $+$ \\
		$\ket{3,2}$ & $+$ & $+$ & $-$ & $-$ & $+$ & $+$ & $-$ & $-$ \\
		$\ket{3,4}$ & $+$ & $+$ & $-$ & $-$ & $+$ & $+$ & $-$ & $-$ \\
		$\ket{4,3}$ & $+$ & $-$ & $-$ & $+$ & $+$ & $-$ & $-$ & $+$ \\
		$\ket{4,5}$ & $-$ & $+$ & $+$ & $-$ & $+$ & $-$ & $-$ & $+$ \\
		$\ket{5,4}$ & $+$ & $-$ & $+$ & $+$ & $-$ & $-$ & $+$ & $-$ \\
		$\ket{5,6}$ & $-$ & $+$ & $+$ & $+$ & $-$ & $-$ & $-$ & $+$ \\
		$\ket{6,5}$ & $-$ & $+$ & $+$ & $+$ & $-$ & $-$ & $-$ & $+$ \\
		$\ket{6,1}$ & $+$ & $-$ & $+$ & $+$ & $-$ & $-$ & $+$ & $-$ \\
		\noalign{\smallskip}\hline
	\end{tabular}
	\end{center}
\end{table}

We now prove that this behavior, where the system only evolves by sign flips, occurs in general for cycles of $N$ vertices with any configuration of marked vertices. First note that in the cycle, each vertex has two neighbors, so in the bipartite double cover, each vertex also has two incident edges. Initially, all edges have the same amplitude. At marked vertices, the reflection operators $R_a'$ and $R_b'$ flip the signs of the incident edges, whereas at unmarked vertices, they invert about the average. For this inversion, the two edges incident to an unmarked vertex either have the same sign or opposite sign. If they have the same sign, the inversion does nothing, and if they have opposite signs, the inversion flips both. So whether edges are incident to marked or unmarked vertices, the quantum walk can only evolve by sign flips.

This proves that any search problem on the 1D cycle is an exceptional configuration for search by Szegedy's quantum walk (or its equivalent coined quantum walk \cite{Wong26}). This encompasses a result by Santos and Portugal \cite{Santos2010a}, who showed through more elaborate analysis that a marked cluster of vertices on the 1D cycle only evolves by sign flips. Our result is more general, since the marked vertices do not need to be clustered together.

This implies that the system remains in an equal probability distribution for all time, which means each vertex of the cycle is equally likely to be measured. Thus, the quantum walk is equivalent to randomly guessing for a marked vertex, so if there are $k$ marked vertices, the probability of guessing one of them is $k/N$, and the expected number of guesses (or repetitions of the algorithm) in order to find one is $O(N/k)$.


\section{Hitting and Mixing Times}

We now use this observation to construct a search problem where the gap between the classical walk's hitting time and the quantum walk's runtime is arbitrary large.

\begin{figure}
\begin{center}
	\includegraphics{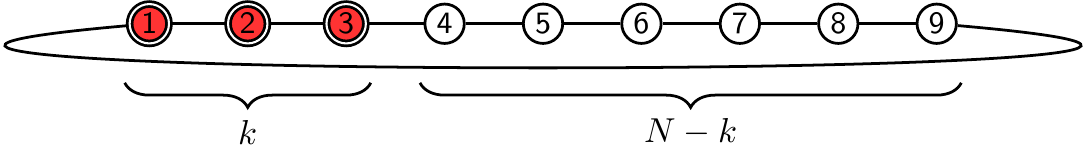}
	\caption{\label{fig:line_N9} A cycle of $N = 9$ vertices, $k = \sqrt{N} = 3$ of which are marked in a cluster.}
\end{center}
\end{figure}

Consider a cycle of $N$ vertices, where $k = o(N)$ of them are marked and contiguous, as in Fig.~\ref{fig:line_N9}. Since Szegedy's quantum walk only evolves by sign flips for any configuration of marked vertices on the cycle, it is equivalent to classically guessing for a marked vertex, so its runtime is $O(N/k)$.

For the classical random walk, let us find its hitting time $HT$. The walker has uniform probability of starting at any of the vertices. If it starts at any of the $k$ marked vertices, then it takes $0$ steps to reach a marked vertex since it started at one. On the other hand, if it starts at one of the $N-k$ unmarked vertices, then it takes some time to reach a marked vertex at its ends. But this unmarked segment is equivalent to searching a cycle of length $L = N - k + 1$ for a marked vertex, and from the Appendix, this has a hitting time of $\mathrm{\Theta}(L^2) = \mathrm{\Theta}(N^2)$. Then
\[ HT = \frac{1}{N} \left[ k \cdot 0 + (N-k) \cdot \mathrm{\Theta}(N^2) \right] = \mathrm{\Theta}(N^2). \]

So while the quantum walk searches in $O(N/k)$ steps, the classical random walk takes $\mathrm{\Theta}(N^2)$. This allows us to construct arbitrary separations between the runtime of the quantum walk and classical random walk's hitting time. For example, when $k = \sqrt{N}$, the quantum walk's runtime is $O(\sqrt{N})$ compared to the classical random walk's hitting time of $O(N^2)$, which is a quartic separation. Or when $k = N/\log N$, we get an exponential separation. If $k = N/\log(\log N)$, we get a doubly exponential separation, and so forth. This seems to be the first example of a greater-than-quadratic separation between the runtime of a quantum walk and the hitting time of a classical random walk. Note that a cubic speedup was shown in \cite{Wong11}, but it requires that the quantum and classical walks each take multiple walk steps per oracle query, so our present result would be the first in the usual setting.

One may object to this arbitrary speedup because the quantum walk is allowed to restart and hence sample from a uniform distribution, whereas the classical random walk is not allowed to. That is, the hitting time is defined such that the random walker jumps around until a marked vertex is found, which makes it differ from repeated random sampling. In this regard, the arbitrary speedup is moreso about the computational power of random sampling versus randomly walking until a marked vertex is found. This suggests that the usual comparison of a quantum walk to a classical random walk's hitting time may not be the most appropriate comparison.

To reconcile this, we can include the time it takes to prepare the initial state in the total cost of the algorithm \cite{Magniez2011}. Typically, the initial state is the uniform superposition or distribution, although a different starting state for the continuous-time quantum walk was explored in \cite{Wong19}. The uniform state expresses our lack of information at the beginning of the computation---any vertex could be marked, so we guess all of them equally.

For a universal circuit-based quantum computer, one can prepare the initial uniform superposition by initializing each of the qubits to $\ket{0}$ and then applying the Hadamard gate to each, and this procedure is often described when introducing Grover's algorithm \cite{NielsenChuang2000}. Restricted to local operations, one can prepare the uniform state on the cycle of $N$ vertices from an initially localized state using $O(N)$ local operations, since that it how many steps it takes to move amplitude to every vertex of the cycle. The same $O(N)$ cost holds for classical local operators preparing the uniform distribution. If one specifically applies a walk as opposed to general local operations, then the time to reach a roughly uniform state from an initially localized state is the \emph{mixing time}, and the quantum walk's is $O(N \log N)$ compared to the classical random walk's $O(N^2)$ \cite{Aharonov2001}. Finally, for quantum search algorithms where the success probability builds up at a marked vertex with high probability, one can run the search algorithm backwards from a known marked vertex to produce the initial uniform state \cite{CG2004}.

Of these choices, since we are investigating search by quantum and random walk, we choose the option where the initial state must also be prepared using a walk. Multiplying the quantum walk's mixing time of $O(N \log N)$ for the cycle with its expected $O(N/k)$ repetitions, we get an overall runtime of $O(N^2 \log N / k)$. Since $k = o(N)$, this runtime scales greater than $N \log N$. The classical algorithm is only run once, so we add its mixing time of $O(N^2)$ to its hitting time of $O(N^2)$ for an overall runtime of $O(N^2)$. Thus, the quantum speedup is now just short of quadratic instead of arbitrarily large. Note that using the classical random walk to randomly sample would be foolish since the steps to prepare the uniform state times the expected number of samples is $O(N^3/k)$, which scales greater than $N^2$ since $k = o(N)$.

This reveals that the hitting time alone is not always the appropriate comparison between the quantum walk and the classical random walk, and the mixing time to prepare the initial state is an important factor.

Of course, if we only count the query complexity of the search algorithms, as is typically done, then the speedup between the quantum walk and the hitting time is indeed arbitrarily large since preparing the initial state requires no queries to the oracle.


\section{Higher-Dimensional Generalizations}

\begin{figure}
\begin{center}
	\subfloat[]{
		\includegraphics{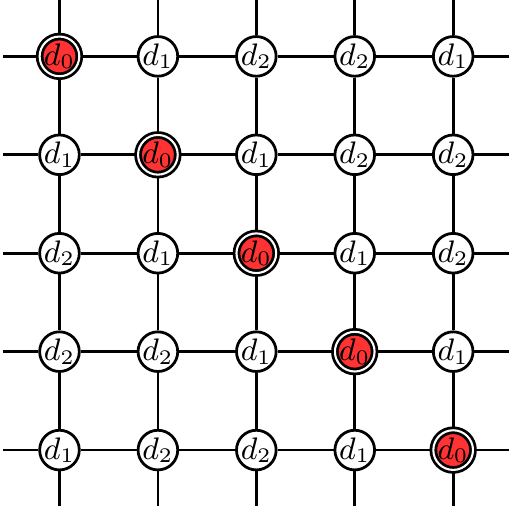}
		\label{fig:grid5_subspace}
	} \quad \quad
	\subfloat[]{
		\includegraphics{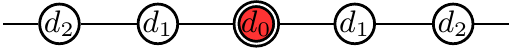}
		\label{fig:grid5_subspace_line}
	}
	\caption{\label{fig:grid5} (a) Periodic square lattice of $N = 5 \times 5$ vertices with a marked diagonal, denoted by double circles and colored red, and (b) its reduction to a 1D search problem. Identically evolving vertices are identically labeled.}
\end{center}
\end{figure}

We end by generalizing our results to higher dimensions, beginning with the 2D grid with a marked diagonal from Fig.~\ref{fig:grid3_diagonal}. Recall that \cite{AR2008} showed this to be an exceptional configuration that only evolves by sign flips. We reproduce their result by simply observing that the 2D grid with a marked diagonal can be reduced to a 1D cycle with a single marked vertex, as shown in Fig.~\ref{fig:grid5}. Since we proved that any configuration of marked vertices on the 1D cycle is exceptional, the same must be true for the 2D grid with a marked diagonal. Then the system stays in a uniform probability distribution, so it is equivalent to classically guessing for a marked vertex. If the grid has $N = \sqrt{N} \times \sqrt{N}$ vertices, the marked diagonal contains $\sqrt{N}$ of them, and the probability of randomly guessing one is $1/\sqrt{N}$. So we expect to make $O(\sqrt{N})$ guesses in order to find a marked vertex. The mixing time for a quantum walk on the 2D grid is $O(\sqrt{N \log N})$, whereas a classical random walk's is $O(N)$ \cite{Marquezino2010}. So the overall runtime of the quantum walk with preparing the initial state and repeating the algorithm to sample for a marked vertex is $O(N \sqrt{\log N})$. If we instead prepare the initial state using local $O(\sqrt{N})$ transformations \cite{AKR2005}, then the total runtime is $O(N)$.

Now let us compare this to the hitting time of a classical random walk. In the Appendix, we prove that the hitting time of a cycle of $L$ vertices with a single marked vertex is exactly $(L^2-1)/6$. In reducing a 2D grid with a marked diagonal to a cycle with a single marked vertex, we have $L = \sqrt{N}$, so the hitting time is $O(N)$, which is the same order as the mixing time. This indicates that the quantum walk does not yield a speedup when searching the 2D grid for a marked diagonal.

Finally, any other search problem that can be reduced to the cycle is an exceptional configuration. For example, search on the 3D periodic cubic lattice with an appropriately marked diagonal reduces to search on the cycle, and so this is also an exceptional configuration. Higher dimensional generalizations also apply. Thus, there is a whole family of exceptional configurations for search by Szegedy's quantum walk.


\section{Conclusion}

Quantum walks typically search for marked vertices in graphs by accumulating amplitude at them. The 2D periodic square lattice with a marked diagonal, however, is an exception to this, only evolving by its amplitudes acquiring minus signs. Hence, it is equivalent to classically guessing and checking for a marked vertex. We proved that the same behavior occurs for Szegedy's quantum walk, or its equivalent coined quantum walk \cite{Wong26}, for any configuration of marked vertices on the 1D cycle. Central to our proof was an observation that the reflection operators in Szegedy's quantum walk performs an inversion about an average. This also holds for any search problem that reduces to a 1D cycle, so there is a whole family of exceptional configurations for which quantum walks do not spread.

Utilizing this phenomenon, we constructed a search problem on the cycle with a contiguous cluster of marked vertices where the quantum walk's random sampling is arbitrarily faster than the classical random walk's hitting time. This only counts the query complexity or assumes that the initial uniform state is prepared for free. In this regard, this result is moreso about the computational power of random sampling versus randomly walking until a marked vertex is found. If the initial state must be constructed using a walk, however, then the quantum walk now obtains a nearly quadratic speedup over the classical random walk.

An open question is whether there are exceptional configurations outside of the family we have identified where the system only evolves by sign flips. Another open question is whether the 2D grid with a marked diagonal can be searched more quickly than $O(\sqrt{N})$. Since it takes that many steps to initialize the uniform superposition using local operations, perhaps an alternative initial state will need to be used.


\begin{acknowledgements}
	T.W.~thanks the quantum computing group at the University of Texas at Austin for useful discussions.
	T.W.~was supported by the U.S.~Department of Defense Vannevar Bush Faculty Fellowship of Scott Aaronson.
	R.S.~was supported by the RAQUEL (Grant	Agreement No. 323970) project, and the ERC Advanced Grant MQC. 
\end{acknowledgements}


\appendix
\section*{Appendix: Hitting Time on the Cycle}

In this appendix, we derive the hitting time $HT$ of a classical random walk on the cycle of $L$ vertices with a single marked vertex and show that it is $HT = (L^2 - 1)/6$.

Say the random walker is currently at distance $i$ from the marked vertex. Then let $h_i$ be the expected number of steps needed to move \emph{one step closer} to the marked vertex from this position. Trivially, $h_0 = 0$, so let us find an expression for $h_i$ with $i > 0$, depending on if $L$ is even or odd.

If $L$ is even, then $i$ is at most $L/2$, and
\[ h_{L/2} = 1 \]
since the walker can only move closer to the marked vertex. We also have in general for $i < L/2$ that
\[ h_i = \frac{1}{2} (1) + \frac{1}{2} \left( 1 + h_{i+1} + h_i \right), \]
where first term comes from immediately moving closer to the marked vertex, and the second term comes from moving a step away first. Solving this for $h_i$,
\[ h_i = 2 + h_{i+1}. \]
By recursion,
\[ h_i = 2 \left( \frac{L}{2} - i \right) + h_{L/2} = L - 2i + 1. \]

If $L$ is odd, then $i$ is at most $(L-1)/2$, and there are two vertices at this distance from the marked vertex. Then
\[ h_{(L-1)/2} = \frac{1}{2} (1) + \frac{1}{2} \left( 1 + h_{(L-1)/2} \right). \]
The first term comes from immediately moving closer to the marked vertex, and the second term comes from moving to the other vertex that is a distance of $(L-1)/2$ away from the marked vertex. Solving for $h_{(L-1)/2}$,
\[ h_{(L-1)/2} = 2. \]
Another way to arrive at this result is noting that
\[ h_{(L-1)/2} = \frac{1}{2} (1) + \frac{1}{4} (2) + \frac{1}{8} (3) + \frac{1}{16} (4) + \dots = \sum_{i=1}^{\infty} \frac{i}{2^i} = 2, \]
where the infinite series was shown to equal $2$ by Oresme in the mid-fourteenth century \cite{Horadam1974}. When $i < (L-1)/2$, we again have that $h_i = \frac{1}{2} (1) + \frac{1}{2} \left( 1 + h_{i+1} + h_i \right)$, and applying this recursively yields
\[ h_i = 2 \left( \frac{L-1}{2} - i \right) + h_{(L-1)/2} = L - 2i + 1. \]
So whether $L$ is even or odd, $h_i$ takes the same value.

Now let $H_i$ denote the expected time to reach the marked vertex if the walker starts a distance $i$ from it. Trivially, $H_0 = 0$. Using the above formula for $h_i$, we find $H_i$ when $i > 0$:
\[ H_i = \sum_{j=1}^i h_j = \sum_{j=1}^i \left[ 2\left(\frac{L}{2}-i\right) + 1 \right] = i(L+1) - 2 \sum_{j=1}^i i = i(L+1) - i(i+1) = i(L-i). \]
An alternative approach to these results so far is given in~\cite{Santos2010b}.
 
Finally, we find the hitting time $HT$ of the random walk, assuming that the walker begins in a uniform distribution over the vertices. If $L$ is even,
\begin{align*}
	HT 
	&= \frac{1}{L} \left[ H_0 + 2 \sum_{i=1}^{L/2-1} H_i + H_{L/2} \right] = \frac{1}{L} \left[ 0 + 2 \sum_{i=1}^{L/2-1} i(L-i) + \frac{L}{2} \left( L - \frac{L}{2} \right) \right] \\
	&= \frac{1}{L} \left[ \frac{L^2}{4} + 2 \left( L \sum_{i=1}^{L/2-1} i - \sum_{i=1}^{L/2-1} i^2 \right) \right].
\end{align*}
Using
\[ \sum_{i=1}^n i = \frac{n(n+1)}{2}, \quad \sum_{i=1}^n i^2 = \frac{n(n+1)(2n+1)}{6}, \]
we get
\[ HT = \frac{L^2 - 1}{6}. \]
If $L$ is odd, the hitting time is
\begin{align*}
	HT 
	&= \frac{1}{L} \left[ H_0 + 2 \sum_{i=1}^{(L-1)/2} H_i \right] = \frac{1}{L} \left[ 0 + 2 \sum_{i=1}^{(L-1)/2} i(L-i) \right] \\
	&= \frac{2}{L} \left( L \sum_{i=1}^{(L-1)/2} i - \sum_{i=1}^{(L-1)/2} i^2 \right) = \frac{L^2 - 1}{6}.
\end{align*}
So whether the cycle is even or odd length, its hitting time is $(L^2-1)/6$.


\bibliographystyle{qinp}
\bibliography{refs}

\end{document}